\documentclass[twoside,leqno]{article}

\pdfoutput=1

\usepackage{RR}
\usepackage[a4paper]{geometry}
\usepackage{a4wide}

\usepackage[english]{babel}
\usepackage[colorlinks=true, citecolor=blue, linkcolor=blue]{hyperref}
\usepackage[numbers]{natbib}

\usepackage{caption}
\usepackage[utf8]{inputenc}
\usepackage[OT1]{fontenc} 

\usepackage{amssymb, amsmath, amsthm, mathtools}
\usepackage{amsfonts, bbold, dsfont}

\usepackage{listings, lstcoq}
\newcommand{\code}[1]{\texttt{#1}}

\usepackage[framemethod=tikz]{mdframed}

\usepackage{tikz}
\usetikzlibrary{arrows.meta,calc,patterns,patterns.meta,%
  positioning,shadings,shapes}

\theoremstyle{plain}
\mdfdefinestyle{thsty}{linecolor=black, roundcorner=5pt, linewidth=1pt}
\newmdtheoremenv[style=thsty]{TextbookTheorem}{Textbook Theorem}
\mdtheorem[style=thsty]{Theorem}{Theorem}
\mdtheorem[style=thsty]{Lemma}[Theorem]{Lemma}
\mdtheorem[style=thsty]{Definition}[Theorem]{Definition}

\newcommand{\AuthorsList}{%
  S. Boldo, F. Clément, V. Martin, M. Mayero, H. Mouhcine}

\newcommand{\LMF}{%
  Université Paris-Saclay, CNRS, ENS Paris-Saclay, Inria,
  Laboratoire Méthodes Formelles,
  91190, Gif-sur-Yvette, France.}
\newcommand{\SERENA}{%
  a. Inria, 2 rue Simone Iff, 75589 Paris, France.\protect\\
  b. CERMICS, École des Ponts, 77455 Marne-la-Vallée, France.}
\newcommand{\LMAC}{%
  Université de technologie de Compiègne, LMAC (Laboratory of
  Applied Mathematics of Compiègne), CS 60319,
  60203 Compiègne Cedex, France.}
\newcommand{\LIPN}{%
  LIPN, Université Paris 13 - USPN, CNRS UMR 7030,
  Villetaneuse, F-93430, France.}

\newcommand{\Funding}{%
  This work was partly supported by the European Research Council (ERC) under
  the European Union's Horizon 2020 Research and Innovation Programme – Grant
  Agreement n$^\circ$810367.
}

\newcommand{\Title}{%
  Lebesgue Induction and Tonelli's Theorem in Coq}
\newcommand{\Titre}{%
  Induction de Lebesgue et théorème de Tonelli en Coq}

\newcommand{\Abstract}{%
  Lebesgue integration is a well-known mathematical tool, used for instance in
  probability theory, real analysis, and numerical mathematics.
  Thus its formalization in a proof assistant is to be designed to fit
  different goals and projects.
  Once Lebesgue integral is formally defined and the first lemmas are proved,
  the question of the convenience of the formalization naturally arises.
  To check it, a useful extension is the Tonelli theorem, stating that the
  (double) integral of a nonnegative measurable function of two variables can
  be computed by iterated integrals, and allowing to switch the order of
  integration.
  Therefore, we need to define and prove results on product spaces, hoping that
  they can easily derive from the existing ones on a single space.
  This article describes the formal definition and proof in {\Coq} of product
  $\sigma$-algebras, product measures and their uniqueness, the construction of
  iterated integrals, up to the Tonelli theorem.
  We also advertise the \emph{Lebesgue induction principle} provided by an
  inductive type for {\nonnegative} measurable functions.
}
\newcommand{\Resume}{%
  L'intégrale de Lebesgue est un outil mathématique bien connu, utilisé par
  exemple en théorie des probabilités, en analyse réelle et pour les
  mathématiques appliquées.
  Sa formalisation dans un assistant de preuve doit donc être conçue pour
  s'adapter des buts et des projets différents.
  Une fois que l'intégrale de Lebesgue est définie formellement et que les
  premiers lemmes sont prouvés, il se pose naturellement la question de la
  commodité d'usage de la formalisation.
  Pour la contrôler, le théorème de Tonelli est une extension utile.
  Ce dernier établi que l'intégrale (double) d'une fonction mesurable positive
  de deux variables peut être calculée par des intégrales itérées et que l'on
  peut intervertir l'ordre d'intégration.
  Nous devons donc définir et prouver des résultats sur les espaces produits,
  en espérant qu'ils peuvent facilement découler des résultats existants sur un
  espace simple.
  Cet article décrit la définition formelle et la preuve en {\Coq} des tribus
  produits, des l'existence et l'unicité des mesures produits, de la
  construction des intégrales itérées, jusqu'au théorème de Tonelli.
  Nous annonçons également le \emph{principe d'induction de Lebesgue}, qui est
  obtenu à partir d'un type inductif pour les fonctions mesurables positives.
}

\newcommand{\Keywords}{%
  Formal proof,
  Coq,
  Measure theory,
  Lebesgue integration,
  Tonelli theorem
}
\newcommand{\Motscles}{%
  Preuve formelle,
  Coq,
  Théorie de la mesure,
  Intégrale de Lebesgue,
  Théorème de Tonelli
}

\newcommand{\myskip}{\bigskip} 

\newcommand{\eg}{e.g.}
\newcommand{\ie}{i.e.}

\newcommand{\byproduct}{by\-product}

\newcommand{\nonempty}{non\-empty}

\newcommand{\nondecreasing}{non\-decreasing}

\newcommand{\nonnegativ}{non\-negativ}
\newcommand{\nonnegative}{{\nonnegativ}e}
\newcommand{\nonnegativity}{{\nonnegativ}ity}

\newcommand{\monotony}{monotonicity} 

\newcommand{\soft}[1]{\textsf{#1}}

\newcommand{\Clang}{\soft{C}}
\newcommand{\Coq}{\soft{Coq}}
  \newcommand{\Coquelicot}{\soft{Coquelicot}}

\newcommand{\Cpp}{\soft{C++}}
  
  \newcommand{\FreeFEM}{\soft{FreeFEM++}}
  \newcommand{\XLiFE}{\soft{XLiFE++}}

\newcommand{\HOLLight}{\soft{HOL Light}}

\newcommand{\Isabelle}{\soft{Isabelle}}
  \newcommand{\IsabelleHOL}{\soft{Isabelle/HOL}}
\newcommand{\Lean}{\soft{Lean}}
\newcommand{\Mizar}{\soft{Mizar}}
\newcommand{\PVS}{\soft{PVS}}
  \newcommand{\PVSNASA}{\soft{PVS-NASA}}

\renewcommand{\leq}{\leqslant}
\renewcommand{\subset}{\subseteq}
\renewcommand{\emptyset}{\varnothing}

\newcommand{\fhi}{\varphi}

\newcommand{\N}{\mathbb{N}}

\newcommand{\R}{\mathbb{R}}
\newcommand{\Rplus}{\R_+}
\newcommand{\Rbar}{\overline{\R}}
\newcommand{\Rbarplus}{\Rbar_+}
\newcommand{\C}{\mathbb{C}}

\newcommand{\calIF}{\mathcal{IF}}
\newcommand{\calM}{\mathcal{M}}
\newcommand{\calMplus}{\calM_+}

\newcommand{\calS}{\mathcal{S}}
\newcommand{\calSF}{\mathcal{SF}}
\newcommand{\calSFplus}{\calSF_+}
\newcommand{\calT}{\mathcal{T}}

\newcommand{\tm}{\tilde{m}}

\newcommand{\charac}[1]{{\mathds{1}}_{#1}}

\newcommand{\Implies}{\Rightarrow}

\newcommand{\AND}{\quad\mbox{and}\quad}
\newcommand{\eqdef}{\stackrel{\mathrm{def.}}{=}}

\newcommand{\preimage}[2]{#1^{-1}(\{#2\})}

\newcommand{\oltimes}{\overline{\times}}
\newcommand{\olSigma}{\overline{\Sigma}}
\newcommand{\imgmeas}{\text{\raisebox{.125em}{{\tiny \#}}}}

\newcommand{\lem}[1]{#1 lemma} 
\newcommand{\thm}[1]{#1 theorem} 
\newcommand{\thms}[1]{\thm{#1}s}
\newcommand{\BL}{Beppo Levi}

\newcommand{\BLmct}{\thm{{\BL} (monotone convergence)}}
\newcommand{\BeppoLeviMonotConvTh}{\BLmct}

\newcommand{\Fat}{Fatou}
\newcommand{\Fl}{\lem{{\Fat}'s}}
\newcommand{\FatouLem}{\Fl}
\newcommand{\Fub}{Fubini}
\newcommand{\Ft}{\thm{\Fub}}
\newcommand{\FubiniTh}{\Ft}

\newcommand{\LM}{Lax--Milgram}
\newcommand{\LMt}{\thm{\LM}}
\newcommand{\LaxMilgramTh}{\LMt}
\newcommand{\LIP}{Lebesgue induction principle}
\newcommand{\LebesgueInductionPrinciple}{\LIP}
\newcommand{\mct}{\thm{monotone class}}
\newcommand{\MonotClassTh}{\mct}
\newcommand{\Ton}{Tonelli}
\newcommand{\Tont}{\thm{\Ton}}
\newcommand{\TonelliTh}{\Tont}
\newcommand{\TonelliAndFubiniThs}{\thms{{\Ton} and {\Fub}}}

\usepackage{color}

\definecolor{tanIII}{RGB}{205,135,63}
\definecolor{yellowII}{RGB}{238,238,0}
\definecolor{darkOliveGreenIII}{RGB}{162,205,90}
\definecolor{turquoiseII}{RGB}{0,229,238}

\definecolor{darkred}{rgb}{0.8,0.2,0.2}
\definecolor{darkgreen}{rgb}{0.2,0.8,0.2}
\definecolor{darkblue}{rgb}{0.2,0.2,0.8}
\definecolor{lightred}{RGB}{255,225,225}
\definecolor{lightgreen}{RGB}{200,255,200}
\definecolor{lightblue}{RGB}{225,225,255}

\definecolor{darkgray}{gray}{0.45}
\definecolor{gray}{gray}{0.60}
\definecolor{lightgray}{gray}{0.75}
\definecolor{lightlightgray}{gray}{0.90}

\def\iscolor{1}  
\if\iscolor1
\newcommand{\colorA}{tanIII}
\newcommand{\colorB}{yellowII}
\newcommand{\colorC}{darkOliveGreenIII}
\newcommand{\colorD}{turquoiseII}
\newcommand{\secAColorName}{\colorbox{\colorA}{brown}}
\newcommand{\secBColorName}{\colorbox{\colorB}{yellow}}
\newcommand{\secCColorName}{\colorbox{\colorC}{green}}
\newcommand{\secDColorName}{\colorbox{\colorD}{blue}}
\else
\newcommand{\colorA}{darkgray}
\newcommand{\colorB}{gray}
\newcommand{\colorC}{lightgray}
\newcommand{\colorD}{lightlightgray}
\newcommand{\secAColorName}{\colorbox{\colorA}{dark gray}}
\newcommand{\secBColorName}{\colorbox{\colorB}{gray}}
\newcommand{\secCColorName}{\colorbox{\colorC}{light gray}}
\newcommand{\secDColorName}{\colorbox{\colorD}{lighter gray}}
\fi

\newcommand{\st}{{\,|\,}}
\newcommand{\mysubsetAinProd}{~$A\!\subset\!X_1\times X_2$}
\newcommand{\mypointa}{}
\newcommand{\mypointb}{ a point}
\newcommand{\mydisplaymaths}[1]{\[#1.\]}

\RRdate{February 2022}

\RRauthor{%
  Sylvie Boldo\thanks[lmf]{{\LMF}
    \texttt{\{sylvie.boldo,houda.mouhcine\}@inria.fr}}
  \and François Clément\thanks[serena]{{\SERENA}
    \texttt{francois.clement@inria.fr}}
  \and Vincent Martin\thanks{{\LMAC} \texttt{vincent.martin@utc.fr}}
  \and Micaela Mayero\thanks[lipn]{{\LIPN}\goodbreak
    \texttt{mayero@lipn.univ-paris13.fr}}
  \and Houda Mouhcine\thanksref{lmf}\thanksref{lipn}\thanksref{serena}
}
\authorhead{\AuthorsList}

\RRtitle{\Titre}
\RRetitle{\Title}
\titlehead{\Title}

\RRnote{\Funding}

\RRresume{\Resume}
\RRabstract{\Abstract}

\RRmotcle{\Motscles}
\RRkeyword{\Keywords}

\RRprojets{Toccata and Serena}

\RCSaclay

\begin{document}

\RRNo{9457}
\makeRR

\section{Introduction}
\label{sec:Introd}

This work deals with the {\Coq}\footnote{\url{https://coq.inria.fr/}}
formalization of the {\LebesgueInductionPrinciple} and the {\TonelliTh} as a
direct continuation of a previous work~\cite{BCF21}.
Our long term objective is to formally prove in {\Coq} scientific computing
programs and the correctness of parts of a {\Cpp} library, such as
{\FreeFEM}\footnote{\url{https://freefem.org/}} or
{\XLiFE},\footnote{\url{https://uma.ensta-paris.fr/soft/XLiFE++/}} that
implements the Finite Element Method (FEM), a widely used method for
numerically solving Partial Differential Equations (PDEs) arising in different
domains like engineering and mathematical modeling.
With this work, we carry on with our goal: to provide a {\Coq} library usable
by numerician people.
It started with the first development of a real numbers library~\cite{May01},
and then by the first complete experimentation of the formalization and proof
of a numerical program, a small {\Clang} program for the approximated
resolution of the wave equation~\cite{BCF13}.
More recently, the {\LaxMilgramTh}~\cite{BCF17} (for the resolution of a class
of PDEs), then Lebesgue integration of {\nonnegative} measurable functions, the
{\BeppoLeviMonotConvTh} and {\FatouLem}~\cite{BCF21}, Bochner
integration~\cite{BCL22} (a generalization of Lebesgue integration for
functions taking their values in a Banach space), and the construction of the
Lebesgue measure%
\footnote{\url{https://lipn.univ-paris13.fr/coq-num-analysis/tree/Tonelli.1.0/Lebesgue/measure_R.v}}
(yet unpublished) have also been formalized.

The proof of the {\TonelliTh} is the next step.
But, as a side result, it also allows us to validate our previous developments
and in particular our definitions and results about the Lebesgue integral.
The validation of a usable development is indeed important.
It should allow us to carry on by confirming or not the choices of
formalization.
For example, as we work in {\Coq}, the question of using classical or
intuitionistic real analysis is a valid question.
As explained in~\cite{BCF17} and~\cite{BCF21}, our view on the question
has evolved.
In this work, we make the same choices as in the latter, namely we are
completely classical.

\myskip

The {\LebesgueInductionPrinciple} is a proof technique for properties about
{\nonnegative} measurable functions, and usually involving the integral.
It reflects the three construction steps followed by Henri Lebesgue to build
his integral~\cite{leb:lir:04}.
The property is first established for indicator functions, then for
{\nonnegative} simple functions by checking that the property is compatible
with positive linear operations, and finally for all {\nonnegative} measurable
functions by checking that it is compatible with the supremum.
This technique is an important asset for the proof of the {\TonelliTh}, and we
provide it as a {\byproduct} of an inductive type.

The {\TonelliTh} provides a convenient way to ease the computation of
multiple integrals by stating their equality with iterated integrals, each in
a single dimension.
The {\TonelliTh} applies to {\nonnegative} measurable functions.
A similar result, the {\FubiniTh}, applies to integrable functions with
arbitrary sign, or even taking their values in a Banach space when using the
Bochner integral.
Both theorems can be combined to ease the proof of integrability of the
multi-variable function to integrate.
This article focuses on the case of {\nonnegative} functions, and as usual
in mathematics, we are only interested in the case of two variables.

We aim to the construction of the full formal proof in {\Coq} of the
{\TonelliTh}, stating that the (double) integral of a {\nonnegative} measurable
function of two variables can be computed by iterated integrals, and allowing
to switch the order of integration.
It can be expressed in a mathematical setting as follows.
\pagebreak
\begin{Theorem}[{\Ton}]
  \label{th:tonelli}
  Let $(X_1,\Sigma_1,\mu_1)$ and~$(X_2,\Sigma_2,\mu_2)$ be measure spaces.
  Assume that~$\mu_1$ and~$\mu_2$ are $\sigma$-finite.
  Let $f\in\calMplus(X_1\times X_2,\Sigma_1\otimes\Sigma_2)$.
  Then, we have
  \begin{gather}
    \label{eq:tonelli1}
    \big( \forall x_1 \in X_1,\;
    f_{x_1} \in \calMplus (X_2, \Sigma_2) \big)
    \quad\land\quad
    \int_{X_2} f_{x_1} \, d\mu_2 \in \calMplus (X_1, \Sigma_1),\\
    \label{eq:tonelli2}
    \big( \forall x_2 \in X_2,\;
    f^{x_2} \in \calMplus (X_1, \Sigma_1) \big)
    \quad\land\quad
    \int_{X_1} f^{x_2} \, d\mu_1 \in \calMplus (X_2, \Sigma_2),\\
    \label{eq:tonelli3}
    \int_{X_1\times X_2} f \, d(\mu_1 \otimes \mu_2)
    = \int_{X_1} \left( \int_{X_2} f_{x_1} \, d\mu_2 \right) \, d\mu_1
    = \int_{X_2} \left( \int_{X_1} f^{x_2} \, d\mu_1 \right) \, d\mu_2.
  \end{gather}
\end{Theorem}
The notations in this statement are specified in the remainder of this paper.
Just note that many measures, including the Lebesgue measure, are
$\sigma$-finite (defined in Section~\ref{sec:Prod}), ${\calMplus}$~denotes the
set of {\nonnegative} measurable functions (see Section~\ref{sec:MeasFun}),
and~$f_{x_1}$ and~$f^{x_2}$ are partial applications of~$f$ (see
Section~\ref{sec:SecFun}).
Notice also that the properties~\eqref{eq:tonelli1} and~\eqref{eq:tonelli2}
ensure the existence of all simple integrals, while the existence of the double
integral is granted by the assumption on the function~$f$.

The mathematical definitions and proofs are taken from
textbooks~\cite{mai:m2:14,gh:mip:13,cm:li:21}, and the {\Coq} code is available
at (mainly in files \texttt{Tonelli.v}, \texttt{LInt\_p.v} and \texttt{Mp.v}):
\begin{center}
  {\small \url{https://lipn.univ-paris13.fr/coq-num-analysis/tree/Tonelli.1.0/Lebesgue}\\
  where the tag \texttt{Tonelli.1.0} corresponds to the code of this article
  from {\Coq} $\geq$ 8.12.2.}
\end{center}

\myskip

The {\TonelliTh} is known enough and useful enough to have been formalized
before our work in several proof assistants.
It has been done in {\PVS} in the {\PVSNASA} library%
\footnote{\url{https://github.com/nasa/pvslib/blob/master/measure_integration/fubini_tonelli.pvs}}
by Lester, probably as a follow-up of~\cite{Les07}.
Some Fubini-like results are available in {\HOLLight}~\cite{Har13}.
More recently, the {\TonelliTh} was formalized in {\Mizar} by
Endou~\cite{Endou2019}.

The formalizations nearest to ours are in {\IsabelleHOL} and {\Lean}.
In {\IsabelleHOL}, Hölzl and Heller defined binary and iterated product measure
before the {\FubiniTh}~\cite{HolHel11}.
It cleverly relies on {\Isabelle} type classes and locales.
A more recent work%
\footnote{\url{https://isabelle.in.tum.de/library/HOL/HOL-Analysis/Bochner_Integration.html}}
extends it to the Bochner integral.
In {\Lean}, van Doorn defines products of measures and properties of the
product space towards the {\TonelliAndFubiniThs} in a way very similar to
ours~\cite{VanDoo21} with the same inductive definitions and the same proof
path.
Instead of Lebesgue integral, the {\FubiniTh} is proved with the more generic
Bochner integral.

A very recent (unpublished to our knowledge) work in {\Coq} has been developed
for probability theory.%
\footnote{\url{https://github.com/jtassarotti/coq-proba}}
Many definitions are similar to ours~\cite{BCF21}.
The {\TonelliAndFubiniThs} are proved, but in a quite simpler setting than
ours, as their goal is probability, where the measures are finite.
The $\sigma$-finiteness as above is skipped, and this corresponds in the sequel
to the first parts in proofs of Sections~\ref{sec:MeasSect}
and~\ref{sec:ProdMeas}.

The {\LebesgueInductionPrinciple} is formalized in {\Lean}~\cite{VanDoo21}.
To our knowledge, no formalization is achieved starting from an inductive type.

For a comparison of Lebesgue integral in various proof assistants, we refer the
reader to~\cite{BCF21,VanDoo21}, and we refer to~\cite{BLM16} for a
wider comparison of real analysis in proof assistants.

\myskip

This paper is organized as follows.
Section~\ref{sec:Prereq} gives a brief summary of prerequisites and the main
concepts of measure and integration theories developed in previous works.
The formalization of the {\LebesgueInductionPrinciple} is detailed in
Section~\ref{sec:LIP}.
Section~\ref{sec:Prod} describes the construction of the product measure, while
Section~\ref{sec:Ton} is devoted to the construction of the iterated integrals
and the full proof of the {\TonelliTh}.
Finally, Section~\ref{sec:Concl} concludes and provides hints to future work.

\section{Prerequisites}
\label{sec:Prereq}

Our formalizations and proofs are conducted in {\Coq}.
In this section, we present the necessary prerequisites and libraries for our
developments, from external packages to our own previous work.

\subsection{The {\Coquelicot} Library, $\Rbar$ and Logic}
\label{sec:coquelicot}

The {\Coquelicot}\footnote{\url{http://coquelicot.saclay.inria.fr/}}
library~\cite{BLM15} is a conservative extension of the standard {\Coq} library
of real numbers~\cite{Link_Coq_Ref,May01}.
It provides the formalization of basic results in real analysis for {\Coq}
developments.
Besides the fact that it is a classical library, a salient feature is that it
provides total functions, {\eg} for limit, derivative, and (Rieman) integral.
This is consistent with classical logic, and it means a much simpler and
natural way to write mathematical formulas and theorem statements.
The library also provides a formalization of the extended real numbers
$\Rbar:=\R\cup\{-\infty,+\infty\}$ equipped, among other operations, with
\coqe{Rbar_lub : (Rbar -> Prop) -> Rbar} for the least-upper bound of subsets
of~$\Rbar$, and \coqe{Sup_seq : (nat -> Rbar) -> Rbar} for the supremum of
sequences.

As in the {\Coquelicot} library, we use the full classical logic: total order
on real numbers, propositional and functional extensionality axioms, excluded
middle and choice axioms.

A more detailed description of what we need can be found
in~\cite[Section~2]{BCF21}.

\subsection{Lebesgue Integration Theory}
\label{sec:Lebesgue}

The theory of integration is commonly built upon the measure theory.
The first step defines the measurability of subsets, and then the measure
associates a (possibly infinite) {\nonnegative} number to each measurable
subset.
The second step defines the measurability of functions, and then the integral
associates a (possibly infinite) {\nonnegative} number to each {\nonnegative}
measurable function.
The integral for functions with arbitrary sign is not relevant to the present
work.

This section briefly reviews the main concepts of measure and integration
theories that were presented in~\cite{BCF21} and are needed here.
It includes the notion of generators of $\sigma$-algebra for measurability, and
of adapted sequences to approximate from below measurable functions by simple
functions.

\subsubsection{Measurable Subsets}
\label{sec:MeasSub}

A measurable space~$(X,\Sigma)$ is made of a set~$X$, and the
collection~$\Sigma$ of all its measurable subsets.
The collection~$\Sigma$ is a subset of the power set of~$X$ called
\emph{$\sigma$-algebra}.
It is closed under most subset operations, such as complement, countable union
and countable intersection.
A $\sigma$-algebra can be \emph{generated} as the closure of a smaller
collection of subsets with respect to some of the subset operations.
In our {\Coq} developments, the generators on~\coqe{X : Type} are typically
denoted~\coqe{genX}, and a subset~\coqe{A : X -> Prop} belongs to the
$\sigma$-algebra generated by~\coqe{genX} when the inductive property
\coqe{measurable genX A} holds.

When the set~$X$ has a topological structure, it is convenient to consider its
\emph{Borel $\sigma$-algebra} that is generated by all the open subsets.
The Borel $\sigma$-algebra of~$\Rbar$ can also be generated by the smaller
collection of right closed rays of the form $[a,\infty]$, denoted in {\Coq}
by~\coqe{gen_Rbar}.

Given two measurable spaces~$(X_1,\Sigma_1)$ and~$(X_2,\Sigma_2)$, the
\emph{product $\sigma$-algebra on~$X_1\times X_2$} is the one generated by the
products of measurable subsets of~$X_1$ and~$X_2$.
Some details are provided in Section~\ref{sec:Prod} where it is a major
ingredient.

\subsubsection{Measure}
\label{sec:Meas}

In a measure space~$(X,\Sigma,\mu)$, there is in addition a
\emph{measure}~$\mu$: a function $\Sigma\to\Rbar$ that is {\nonnegative},
homogeneous ($\mu(\emptyset)=0$), and $\sigma$-additive.
This is represented in {\Coq} by a record collecting the support function and
the three constitutive properties.

The properties of \emph{continuity from below} and \emph{from above} are useful
in Section~\ref{sec:Prod}.
For any measure~$\mu$, and for any sequence~$(A_n)_{n\in\N}\in\Sigma$, they
respectively state
\begin{gather}
  \label{eq:cont-below}
  (\forall n \in \N,\; A_n \subset A_{n + 1})
  \Implies
  \mu \left( \bigcup_{n \in \N} A_n \right)
    = \lim_{n \to \infty} \mu (A_n)
    = \sup_{n \in \N} \mu (A_n),\\
  \label{eq:cont-above}
  (\forall n \in \N,\; A_{n + 1} \subset A_n)
  \land
  (\exists n_0 \in \N,\; \mu (A_{n_0}) < \infty)
  \Implies
  \mu \left( \bigcap_{n \in \N} A_n \right)
    = \inf_{n \in \N} \mu (A_n)
\end{gather}
Note that {\monotony} of measures allows to replace the limit of a
{\nondecreasing} sequence by its supremum.
This property of real numbers is repeatedly used in the sequel.

\subsubsection{Measurable Functions}
\label{sec:MeasFun}

Given two measurable spaces~$(X,\Sigma)$ and~$(Y,\calT)$, a function~$f:X\to Y$
is said \emph{measurable} when the preimage of every measurable subset is
measurable:
\begin{lstlisting}
Definition measurable_fun : (X -> Y) -> Prop :=
  fun f => \forall B, measurable genY B -> measurable genX (fun x => B (f x)).
\end{lstlisting}

When $Y:=\Rbar$, and usually~$\calT$ is its Borel $\sigma$-algebra, we may
simply say that the function is \emph{$\Sigma$-measurable}, and we use the
predicate \coqe{measurable_fun_Rbar} corresponding to \coqe{genY := gen_Rbar}.
We denote the \emph{set of {\nonnegative} measurable functions}
by~$\calMplus(X,\Sigma)$.
When there is no possible confusion in the context, we may drop the
``$(X,\Sigma)$'' annotation.
Among other operations, $\calMplus$~is closed under {\nonnegative} scalar
multiplication, addition, and supremum.
In {\Coq}, we use the predicate \coqe{Mplus genX : (X -> Rbar) -> Prop} that
encompasses {\nonnegativity} and measurability, and
\coqe{Mplus_seq genX : (nat -> X -> Rbar) -> Prop} means that all the elements
of a sequence of functions belong to~$\calMplus$.

Two subsets of functions are of major interest for the construction of Lebesgue
integration.
\emph{Simple functions} are functions with range of finite cardinal, and the
\emph{set of {\nonnegative} measurable simple functions} is
denoted~$\calSFplus(X,\Sigma)$.
In {\Coq}, we use the predicate
\coqe{SFplus genX : (X -> Rbar) ->}\linebreak[0]\coqe{Prop},
and any simple function~$f$ is canonically represented by the strictly sorted
list~$\ell$ of its values, $f=\sum_{v\in\ell}v\times\charac{\preimage{f}{v}}$.
Given any function~$f\in\calMplus$, a simple algorithm allows to build an
\emph{adapted sequence for~$f$}, {\ie} a {\nondecreasing} sequence of simple
functions $(\fhi_n)_{n\in\N}\in\calSFplus$ such that
$f=\lim_{n\to\infty}\fhi_n=\sup_{n\in\N}\fhi_n$.
In~\cite{BCF21}, the process, denoted \coqe{mk_adapted_seq}, is obtained
via a fixed-point rounding downwards with a least significant bit of~$-n$
relying on the Flocq library~\cite{BolMel11}.

The \emph{set of measurable indicator functions} is denoted~$\calIF(X,\Sigma)$.
Note that an indicator function~$\charac{A}$ is measurable whenever its support
subset~$A$ belongs to~$\Sigma$.
Simple functions in~$\calSFplus$ are positive linear combinations of indicator
functions in~$\calIF$.

\subsubsection{Lebesgue Integral}
\label{sec:LInt_p}

The construction of the Lebesgue integral in~$\calMplus$ operates in three
steps.
The first stage is to integrate indicator functions in~$\calIF$ by taking the
measure of their support.
Then, the second stage extends the integral to simple functions in~$\calSFplus$
by positive linearity.
And finally, the third stage extends it again to measurable functions
in~$\calMplus$ by taking the supremum.

In the end, the \emph{integral of a function~$f\in\calMplus$} is defined as the
supremum of the integrals of all simple functions in~$\calSFplus$ smaller
than~$f$.
It is formalized in~\cite{BCF21} as
\begin{lstlisting}
Definition LInt_p : (X -> Rbar) -> Rbar :=
  fun f => Rbar_lub (fun z => \exists (phi : X -> R) (Hphi : SF genX phi),
    nonneg phi /\ (\forall x, Rbar_le (phi x) (f x)) /\ LInt_SFp mu phi Hphi = z).
\end{lstlisting}

The proof of the {\TonelliTh} relies on several properties of the integral
in~$\calMplus$, such as {\monotony}, positive linearity, $\sigma$-additivity,
and the {\BeppoLeviMonotConvTh}.
The latter states the compatibility with the supremum: for any {\nondecreasing}
sequence~$(f_n)_{n\in\N}\in\calMplus$, the limit $\lim_{n\to\infty}f_n$ (which
actually equals $\sup_{n\in\N}f_n$) is also in~$\calMplus$, and the
integral-limit exchange formula holds,
$\int\sup_{n\in\N}f_n\,d\mu=\sup_{n\in\N}\int f_n\,d\mu$.

\section{Lebesgue Induction Principle}
\label{sec:LIP}

Let~$(X,\Sigma)$ be a measurable space.
The properties of the function spaces~$\calMplus$, $\calSFplus$ and~$\calIF$
recalled in Section~\ref{sec:MeasFun} suggest we may represent {\nonnegative}
measurable functions by an inductive type.
Indeed, functions in~$\calMplus$ are the supremum of adapted sequences of
{\nonnegative} measurable simple functions, and functions in~$\calSFplus$ are
positive linear combinations of measurable indicator functions in~$\calIF$.
Moreover, the construction of the Lebesgue integral in Section~\ref{sec:LInt_p}
mimics the associated structural induction principle, and the same principle is
a common proof technique for several results in Lebesgue integration theory,
among which the {\TonelliTh} as noted in~\cite{VanDoo21}.

\myskip

In addition to~\coqe{Mplus} recalled in Section~\ref{sec:MeasFun}, we now
define an inductive type:
\begin{lstlisting}
Inductive Mp : (X -> Rbar) -> Prop :=
  | Mp_charac : \forall A, measurable genX A -> Mp (charac A)
  | Mp_scal : \forall a f, 0 <== a -> Mp f -> Mp (fun x => Rbar_mult a (f x))
  | Mp_plus : \forall f g, Mp f -> Mp g -> Mp (fun x => Rbar_plus (f x) (g x))
  | Mp_sup : \forall f, incr_fun_seq f -> (\forall n, Mp (f n)) -> Mp (fun x => Sup_seq (fun n => f n x)).
\end{lstlisting}
where \coqe{incr_fun_seq f} stands for
\coqe{\forall x n, Rbar_le (f n x) (f (S n) x)}.

We also have an inductive type for~$\calSFplus$ denoted by~\coqe{SFp}, whose
constructors are essentially the same as the first three of~\coqe{Mp}.
Several inductive types equivalent to~\coqe{Mp} are defined in order to split
the proof steps, for instance one is built over~\coqe{SFp}.
They are not given here for the sake of simplicity and brevity.

\myskip

The important point is then the correctness of this definition, compared to the
existing one.
The only delicate part is to obtain that simple functions in~$\calSFplus$ can
actually be represented by such an inductive construction, stated in
\coqe{Lemma SFp_correct : \forall f, SFp f <-> SFplus gen f}.

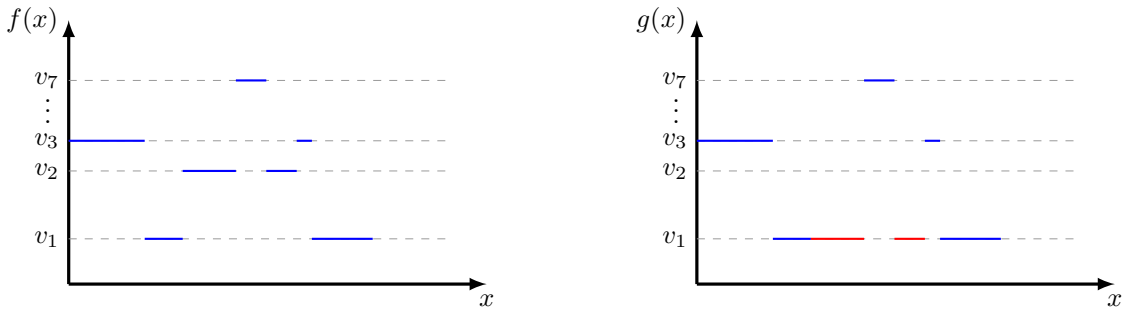
\begin{figure}[bht]
\begin{center}
\begin{tikzpicture}
\draw[very thick,-latex] (0,0) -- (5.5,0) node[below]{$x$};
\draw[very thick,-latex] (0,0) -- (0,3.5) node[left]{$f(x)$};
\draw[help lines,dashed] (0,.6) -- (5,.6);
\node[left] at (0,.6) {$v_1$};
\draw[help lines,dashed] (0,1.5) -- (5,1.5);
\node[left] at (0,1.5) {$v_2$};
\draw[help lines,dashed] (0,1.9) -- (5,1.9);
\node[left] at (0,1.9) {$v_3$};
\node[left] at (-.1,2.4) {$\vdots$};
\draw[help lines,dashed] (0,2.7) -- (5,2.7);
\node[left] at (0,2.7) {$v_7$};
\draw[thick,blue] (0,1.9) -- (1,1.9); 
\draw[thick,blue] (1,.6) -- (1.5,.6); 
\draw[thick,blue] (1.5,1.5) -- (2.2,1.5); 
\draw[thick,blue] (2.2,2.7) -- (2.6,2.7); 
\draw[thick,blue] (2.6,1.5) -- (3,1.5); 
\draw[thick,blue] (3,1.9) -- (3.2,1.9); 
\draw[thick,blue] (3.2,.6) -- (4,.6); 
\end{tikzpicture}
\hfill
\begin{tikzpicture}
\draw[very thick,-latex] (0,0) -- (5.5,0) node[below]{$x$};
\draw[very thick,-latex] (0,0) -- (0,3.5) node[left]{$g(x)$};
\draw[help lines,dashed] (0,.6) -- (5,.6);
\node[left] at (0,.6) {$v_1$};
\draw[help lines,dashed] (0,1.5) -- (5,1.5);
\node[left] at (0,1.5) {$v_2$};
\draw[help lines,dashed] (0,1.9) -- (5,1.9);
\node[left] at (0,1.9) {$v_3$};
\node[left] at (-.1,2.4) {$\vdots$};
\draw[help lines,dashed] (0,2.7) -- (5,2.7);
\node[left] at (0,2.7) {$v_7$};
\draw[thick,blue] (0,1.9) -- (1,1.9); 
\draw[thick,blue] (1,.6) -- (1.5,.6); 
\draw[thick,red] (1.5,.6) -- (2.2,.6); 
\draw[thick,blue] (2.2,2.7) -- (2.6,2.7); 
\draw[thick,red] (2.6,.6) -- (3,.6); 
\draw[thick,blue] (3,1.9) -- (3.2,1.9); 
\draw[thick,blue] (3.2,.6) -- (4,.6); 
\end{tikzpicture}
\end{center}
\caption{%
  Illustration of Lemma \coqe{SF_aux_cons}.
  The value~$v_2$ taken by the simple function~$f$ (on the left) is replaced
  in~$g$ (on the right) by the value~$v_1$ (in red).}
\label{fig:SFauxcons}
\end{figure}

For that, from a simple function represented by a list of values of size~$n+1$,
we need to construct a smaller simple function associated to a sublist of
size~$n$.
The tricky needed result is the following:
\begin{lstlisting}
Lemma SF_aux_cons :
  \forall (f : X -> R) v1 v2 l, nonneg f -> SF_aux genX f (v1 :: v2 :: l) ->
    let g := fun x => f x + (v1 - v2) * charac (fun t => f t = v2) x in
    nonneg g /\ SF_aux genX g (v1 :: l).
\end{lstlisting}
Given~$f\in\calSFplus$ and its associated canonical list~$\ell$, the lemma
builds a new~$g\in\calSFplus$ canonically associated with the list~$\ell$
deprived from some item~$v_2$.
This means that on the {\nonempty} subset~$\preimage{f}{v_2}$, $g$~must take
one of the remaining values, $v_1$~as shown in Figure~\ref{fig:SFauxcons},
which also provides the property~$g\leq f$.

More precisely, let us assume that
$f(x)=\sum_{v\in\{v_1,v_2\}\cup\ell}v\times\charac{\preimage{f}{v}}$.
Then, by setting $g(x):=f(x)+(v_1-v_2)\times\charac{\preimage{f}{v_2}}$, one
has $g(x)=\sum_{v \in\{v_1\}\cup\ell}v\times\charac{\preimage{f}{v}}$.
Thus, $g\in\calSFplus$ with a smaller list of values, and
$f(x)=g(x)+(v_2-v_1)\times\charac{\preimage{f}{v_2}}$ with $v_2-v_1\ge0$.
This is tricky for two reasons.
First, we cannot set~$g$ to zero on~$\preimage{f}{v_2}$ (as it may be a new
value, defeating the point of reducing the size of the value list); thus, the
initial list must contain at least two values.
Second, by proceeding the other way around and setting~$g$ to~$v_2$
on~$\preimage{f}{v_1}$, we cannot write~$f$ as the sum of~$g$ and a
\emph{\nonnegative} value times an indicator function, as needed by the
constructor~\coqe{SFp_scal}, similar to~\coqe{Mp_scal}.

\myskip

Now, we have all the ingredients to check that the definition of~\coqe{Mp} is
satisfactory, that is to say that~\coqe{Mp} represents~$\calMplus$
as~\coqe{Mplus} already does.
This correctness lemma is stated as
\begin{lstlisting}
Lemma Mp_correct : \forall f, Mp genX f <-> Mplus genX f.
\end{lstlisting}
The proof is mainly based on inductions, the construction of adapted sequences
\coqe{mk_adapted_seq} (see Section~\ref{sec:MeasFun}), and the previous lemma.

\myskip

This gives us for free an induction lemma corresponding to the~\coqe{Mp}
inductive:
\begin{lstlisting}
Mp_ind : \forall P : (E -> Rbar) -> Prop,
  (\forall A, measurable gen A -> P (charac A)) ->
  (\forall a f, 0 <== a -> Mp f -> P f -> P (fun x => Rbar_mult a (f x))) ->
  (\forall f g, Mp f -> P f -> Mp g -> P g -> P (fun x => Rbar_plus (f x) (g x))) ->
  (\forall f, incr_fun_seq f -> (\forall n, Mp (f n)) -> (\forall n, P (f n)) -> P (fun x => Sup_seq (fun n => f n x))) ->
  \forall f, Mp f -> P f.
\end{lstlisting}
The corresponding mathematical statement is the following
\begin{Lemma}[\LIP]
Let~$(X,\Sigma)$ be a measurable space.
Let~$P$ be a predicate on functions from~$X$ to~$\Rbar$.
Assume that~$P$ holds on~$\calIF$, and that it is compatible on~$\calMplus$
with positive linear operations and with the supremum of {\nondecreasing}
sequences:
\begin{align}
  \label{eq:LIP1}
  \forall A,\quad &A \in \Sigma \Implies P (\charac{A}),\\
  \label{eq:LIP2}
  \forall a \in \Rplus,\;
  \forall f \in \calMplus,\quad
  &P (f) \Implies P (a f),\\
  \label{eq:LIP3}
  \forall f, g \in \calMplus,\quad
  &P (f) \land P (g) \Implies P (f + g),\\
  \label{eq:LIP4}
  \forall (f_n)_{n \in {\N}} \in \calMplus,\quad
  &(\forall n \in \N,\; f_n \leq f_{n + 1} \land P (f_n)) \Implies
  P \left( \sup_{n \in \N} f_n \right).
\end{align}
Then, $P$~holds on~$\calMplus$.
\end{Lemma}

There are a few alternative statements of the {\LIP}.
For instance, we choose to have~$a$ in~$\R$ and not in~$\Rbar$ in
Equation~\eqref{eq:LIP2}, as it makes an equivalent, but simpler to use lemma.
Moreover, as noted in the {\Lean} source code,%
\footnote{\url{https://leanprover-community.github.io/mathlib_docs/measure_theory/integral/lebesgue.html\#measurable.ennreal_induction}.}
it is possible to sharpen the premises of the constructors.
For instance, it may be sufficient to have in~\eqref{eq:LIP3} simple functions
that do not share the same image value, except~0, or with disjoint supports.

\section{Product Measure on a Product Space}
\label{sec:Prod}

In this section, we build the product measure for the measurable subsets of a
product space.
This allows to integrate numeric functions defined on such a product space in
Section~\ref{sec:Ton}.

Given two measure spaces~$(X_1,\Sigma_1,\mu_1)$ and~$(X_2,\Sigma_2,\mu_2)$, a
\emph{product measure on the measurable space
  $(X_1\times X_2,\Sigma_1\otimes\Sigma_2)$ induced by~$\mu_1$ and~$\mu_2$} is
a measure~$\mu$ defined on the product
$\sigma$-algebra $\Sigma_1\otimes\Sigma_2$ (defined in
Section~\ref{sec:ProdSigAlg}) satisfying the \emph{box property}:
\begin{equation}
  \label{eq:ProdMeasBox}
  \forall A_1 \in \Sigma_1,\;
  \forall A_2 \in \Sigma_2,\quad
  \mu (A_1 \times A_2) = \mu_1 (A_1) \, \mu_2 (A_2).
\end{equation}
To ensure existence and uniqueness of such a product measure, we assume
that~$\mu_1$ and~$\mu_2$ are \emph{$\sigma$-finite} measures, {\ie} that the
full sets~$X_1$ and~$X_2$ are (possibly {\nondecreasing}) unions of subsets of
finite measure (see a detailed definition in Section~\ref{sec:MeasSect}).

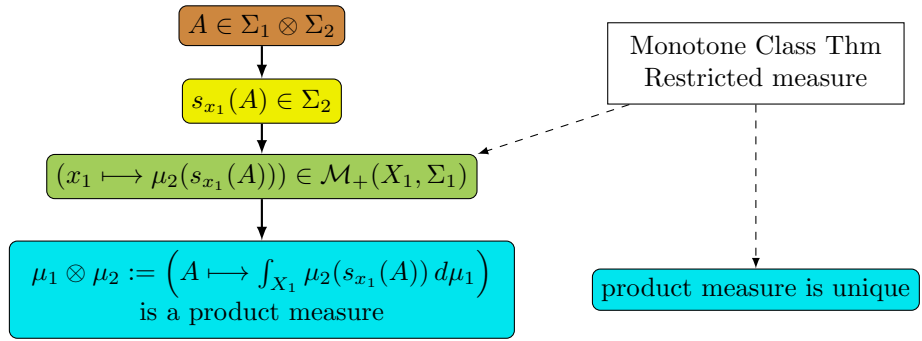
\begin{figure}[htb]
\begin{center}
\begin{tikzpicture}
\node[draw,rounded corners=3pt,fill=\colorA] (S1S2) at (3.5,4.5)
     {$A\in\Sigma_1\otimes\Sigma_2$};
\node[draw,rounded corners=3pt,fill=\colorB] (Sec) at (3.5,3.5)
  {$s_{x_1}(A)\in\Sigma_2$};
\node[draw,rounded corners=3pt,fill=\colorC] (MeasSec) at (3.5,2.5)
  {$\left(x_1\longmapsto\mu_2(s_{x_1}(A))\right)\in\calMplus(X_1,\Sigma_1)$};
\node[draw,rounded corners=3pt,fill=\colorD] (CandProdMeas) at (3.5,1)
  {$\begin{array}{c}
      \mu_1\otimes\mu_2:=
      \left(A\longmapsto\int_{X_1}\mu_2(s_{x_1}(A))\,d\mu_1\right)\\
      \mbox{is a product measure}
    \end{array}$};
\node[draw,rounded corners=3pt,fill=\colorD] (Uniq) at (10,1)
  {product measure is unique};
\draw[->,>=latex,thick] (S1S2) to (Sec);
\draw[->,>=latex,thick] (Sec) to (MeasSec);
\draw[->,>=latex,thick] (MeasSec) to (CandProdMeas);
\node[draw] (MCTh) at (10,4)
  {$\begin{array}{c}
      \mbox{Monotone Class Thm}\\
      \mbox{Restricted measure}
    \end{array}$};
\draw[->,>=latex,dashed] (MCTh) to (MeasSec.north east);
\draw[->,>=latex,dashed] (MCTh) to (Uniq);
\end{tikzpicture}
\end{center}
\caption{%
  Flowchart illustrating the construction of the product measure.\protect\\
  The fill colors refer to sections:
  \ref{sec:ProdSigAlg} in {\secAColorName},
  \ref{sec:Sect} in {\secBColorName},
  \ref{sec:MeasSect} in {\secCColorName},
  and \ref{sec:ProdMeas} in {\secDColorName}.\protect\\
  Dashed lines denote the use of the listed proof arguments, that were
  developed for the present work.
}
\label{fig:prod-meas}
\end{figure}

A candidate product measure is first built in three steps, see
Figure~\ref{fig:prod-meas}.
Firstly, \emph{$X_1$-sections} (or \emph{``vertical'' cuttings}) of subsets are
proved to be $\Sigma_2$-measurable.
Then, the measure of sections is proved to be $\Sigma_1$-measurable.
The candidate is the integral of the measure of sections.
Then, this candidate is proved to be a product measure, and the product measure
is guaranteed to be unique.
The main argument for this construction is the {\MonotClassTh}, whose quite
heavy proof is not detailed here.
It is used twice: for the measurability of the measure of sections, and for the
uniqueness of the product measure.

The definition of the product $\sigma$-algebra is first reviewed in
Section~\ref{sec:ProdSigAlg}.
Then, Section~\ref{sec:Sect} is dedicated to sections, and
Section~\ref{sec:MeasSect} to the measure of sections.
Finally, existence and uniqueness of the product measure is obtained in
Section~\ref{sec:ProdMeas}.

\subsection{Product $\sigma$-algebra}
\label{sec:ProdSigAlg}

Let us detail the notion of product $\sigma$-algebra that was introduced
in~\cite{BCF21}.
Given two measurable spaces~$(X_1,\Sigma_1)$ and~$(X_2,\Sigma_2)$, the
\emph{product $\sigma$-algebra on~$X_1\times X_2$} is the
$\sigma$-algebra~$\Sigma_1\otimes\Sigma_2$ generated by the products of
measurable subsets:
\begin{equation*}
  \Sigma_1 \otimes \Sigma_2 := \mbox{ $\sigma$-algebra generated by }
  \Sigma_1 \oltimes \Sigma_2 :=
  \{ A_1 \times A_2 \;|\; A_1 \in \Sigma_1 \land A_2 \in \Sigma_2 \}
  \ (\subsetneq \Sigma_1 \otimes \Sigma_2).
\end{equation*}
Given generators~\coqe{genX1} and~\coqe{genX2} for~$\Sigma_1$ and~$\Sigma_2$,
the generator~$\Sigma_1\oltimes\Sigma_2$ is denoted in {\Coq} by
\coqe{Product_Sigma_algebra genX1 genX2}.
It is proven in~\cite[Sec. 4.3]{BCF21} that
$\Sigma_1\otimes\Sigma_2$ is also the $\sigma$-algebra generated by
\begin{equation*}
  \{ A_1 \times A_2 \;|\;
  A_1 \in \mathrm{gen}(\Sigma_1) \cup \{ X_1 \} \land
  A_2 \in \mathrm{gen}(\Sigma_2) \cup \{ X_2 \} \}.
\end{equation*}
This smaller generator is denoted in {\Coq} by \coqe{Gen_Product genX1 genX2},
and simply denoted in the sequel by \coqe{genX1xX2}.
Symmetrically, \coqe{genX2xX1} represents \coqe{Gen_Product genX2 genX1}.

\subsection{Section of Subset}
\label{sec:Sect}

\begin{figure}[htb]
\begin{center}
\begin{tikzpicture}
\draw[->,thick] (0,0)--(5,0) node[right]{$X_1$};
\draw[->,thick] (0,0)--(0,3) node[above]{$X_2$};

\coordinate (M) at (4,2.5);
\pgfmathsetmacro{\myslope}{atan2(2,-4.5)}
\draw[fill=gray!30] (M) to[out=\myslope,in=0]
    ++ (-1,0.25) to[out=180,in=90] ++ (-.5,-0.25)
 to[out=-90,in=90] ++ (1,-1) to[out=-90,in=90] ++ (-2.5,-.5)
 to[out=-90,in=180] ++(2,-.5) to[out=0,in=\myslope+180] cycle;
\node at (3.9,2.2) {$A$};

\node[below] at (1.75,0) {$x_1$};
\draw[dashed] (1.75,0)--(1.75,3) node[above]{$s_{x_1}(A)$};
\draw[ultra thick] (1.75,0.45)--(1.75,1.35);

\node[below] at (3.25,0) {$y_1$};
\draw[dashed] (3.25,0)--(3.25,3) node[above]{$s_{y_1}(A)$};
\draw[ultra thick] (3.25,0.54)--(3.25,1.17);
\draw[ultra thick] (3.25,1.88)--(3.25,2.73);
\end{tikzpicture}
\end{center}
\caption{%
  $X_1$-sections of a subset~$A$ of~$X_1\times X_2$ at points~$x_1$ and~$y_1$.}
\label{fig:section}
\end{figure}
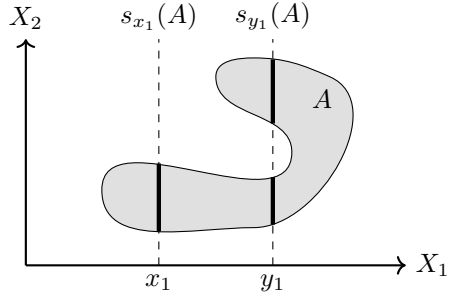

The notion of \emph{section} consists in keeping one of the variables fixed.
Given{\mysubsetAinProd} and{\mypointa}~$x_1\in X_1$, the
\emph{$X_1$-section of~$A$ at~$x_1$} is the subset of~$X_2$ defined by
$s_{x_1}(A):=\{x_2\in X_2\;|\;(x_1,x_2)\in A\}$ (see Figure~\ref{fig:section}).
The {\Coq} translation is straightforward.
\begin{lstlisting}
Definition section : X1 -> (X1 * X2 -> Prop) -> X2 -> Prop := fun x1 A x2 => A (x1, x2).
\end{lstlisting}

Sections commute with most subset operations.
For example, they are compatible with the empty set
($s_{x_1}(\emptyset)=\emptyset$), the complement ($s_{x_1}(A^c)=s_{x_1}(A)^c$),
countable union and intersection, and are monotone.
Sections also satisfy the following box property: for all
subsets~$A_1\subset X_1$, $A_2\subset X_2$, and point~$x_1\in X_1$,
\begin{equation}
  \label{eq:box-section}
  x_1 \in A_1 \Implies s_{x_1} (A_1 \times A_2) = A_2
  \AND
  x_1 \not\in A_1 \Implies s_{x_1} (A_1 \times A_2) = \emptyset.
\end{equation}

Then, we prove that, if a subset~$A$ is $\Sigma_1\otimes\Sigma_2$-measurable,
then its $X_1$-sections at any point in~$X_1$ are $\Sigma_2$-measurable.
As measurability is an inductive type, the proof is a simple induction on the
hypothesis.
\begin{lstlisting}
Lemma section_measurable : \forall A x1, measurable genX1xX2 A -> measurable genX2 (section x1 A).
\end{lstlisting}

\subsection{Measurability of Measure of Section}
\label{sec:MeasSect}

As sections are measurable (see Section~\ref{sec:Sect}), one can take their
measure.
In Section~\ref{sec:ProdMeas}, the product measure is defined as the integral
of the measure of sections, but before that, we have to prove the and
{\nonnegativity} and measurability of these functions.
More precisely, that for all $\Sigma_1\otimes\Sigma_2$-measurable subset~$A$,
the function $(x_1\mapsto\mu_2(s_{x_1}(A)))$ belongs to
$\calMplus(X_1,\Sigma_1)$.

The {\nonnegativity} property directly follows from that of measures.
The proof of measurability goes in two stages.
Firstly when the measure~$\mu_2$ is assumed to be \emph{finite} ({\ie}
when~$\mu_2(X_2)$ is finite), and then in the more general $\sigma$-finite
case.
The first stage is quite high-level, it relies on the {\MonotClassTh}.
The second stage extends the first one by means of restricted measures.

\myskip

After having defined the measure of sections, represented in {\Coq} by the
total function
\begin{lstlisting}
Definition meas_section : (X1 * X2 -> Prop) -> X1 -> Rbar := fun A x1 => muX2 (section x1 A).
\end{lstlisting}
the first stage of the proof is stated in {\Coq} as
\begin{lstlisting}
Lemma meas_section_Mplus_finite :
  \forall A, is_finite_measure muX2 -> measurable genX1xX2 A -> Mplus genX1 (meas_section A).
\end{lstlisting}

Let~$\calS$ be the set of measurable subsets satisfying the property to prove,
\begin{equation*}
  \calS :=
  \left\{ A \in \Sigma_1 \otimes \Sigma_2 \st
  \big( x_1 \longmapsto \mu_2 (s_{x_1} (A)) \big)
  \in \calMplus (X_1, \Sigma_1) \right\}.
\end{equation*}
It suffices to show that $\Sigma_1\otimes\Sigma_2\subset\calS$.
Firstly, $\calS$~is proved to contain the generator
$\olSigma:=\Sigma_1\oltimes\Sigma_2$ of~$\Sigma_1\otimes\Sigma_2$ (see
Section~\ref{sec:ProdSigAlg}).
Then, it is proved to contain the algebra of sets generated by~$\olSigma$
(where an algebra of sets contains the empty set and is closed under complement
and finite union).
Then, $\calS$~is also proved to be a monotone class, {\ie} closed under
monotone countable union and intersection.
This step uses the finiteness assumption on~$\mu_2$, and continuity from below
and from above (see Equations~\eqref{eq:cont-below} and~\eqref{eq:cont-above}).
And finally, we conclude by applying the following {\MonotClassTh} with
\coqe{X := X1 * X2}, \coqe{P := \calS}, and \coqe{genX := \olSigma}.
\begin{lstlisting}
Theorem monotone_class_Prop :
  \forall P : (X -> Prop) -> Prop, is_Monotone_class P ->
    Incl (Algebra genX) P -> Incl (Sigma_algebra genX) P.
\end{lstlisting}
Note that \coqe{Incl} denotes the inclusion for subsets of the power set
of~$X$.

\myskip

In the second stage, the measure~$\mu_2$ is supposed to be $\sigma$-finite.
Thus, there exists a {\nondecreasing} sequence $(B_n)_{n\in\N}\in\Sigma_2$ such
that $X_2=\bigcup_{n\in\N}B_n$, and~$\mu_2(B_n)$ is finite for all~$n\in\N$.
Then, for each~$n\in\N$, the \emph{restricted measure}
\begin{equation*}
  \mu_2^n := (A_2 \in \Sigma_2 \longmapsto \mu_2 (A_2 \cap B_n) \in \Rbarplus)
\end{equation*}
is proved to be a finite measure.
Thus, the previous result applies,
\begin{equation*}
  \forall A \in \Sigma_1 \otimes \Sigma_2,\quad
  (x_1 \longmapsto \mu_2^n (s_{x_1} (A))) \in \calMplus (X_1, \Sigma_1).
\end{equation*}
Moreover, from the properties of sections (see Section~\ref{sec:Sect}) and from
the continuity from below of~$\mu_2$, for all $A\in\Sigma_1\otimes\Sigma_2$
and~$x_1\in X_1$,
\begin{align*}
  \mu_2 (s_{x_1} (A))
  &= \mu_2 \left( s_{x_1} (A) \cap \bigcup_{n \in \N} B_n \right)
  = \mu_2 \left( \bigcup_{n \in \N} s_{x_1} (A) \cap B_n \right)\\
  &= \sup_{n \in \N} \mu_2 \left( s_{x_1} (A) \cap B_n \right)
  = \sup_{n \in \N} \mu_2^n (s_{x_1} (A)).
\end{align*}
Finally, the closedness of~$\calMplus(X_1,\Sigma_1)$ under supremum (see
Section~\ref{sec:MeasFun}) concludes the proof.
Thus, the lemma in the $\sigma$-finite case holds,
\begin{lstlisting}
Lemma meas_section_Mplus_sigma_finite :
  \forall A, is_sigma_finite_measure muX2 -> measurable genX1xX2 A -> Mplus genX1 (meas_section A).
\end{lstlisting}

\myskip

Note that from~\eqref{eq:box-section}, the measure of the section of a box
reads
\begin{equation}
  \label{eq:box-meas-section}
  \forall A_1 \in \Sigma_1,\;
  \forall A_2 \in \Sigma_2,\quad
  (x_1 \longmapsto \mu_2 (s_{x_1} (A_1 \times A_2)))
  = \mu_2 (A_2) \, \charac{A_1}.
\end{equation}

\subsection{Existence and Uniqueness of the Product Measure}
\label{sec:ProdMeas}

As the measures of sections are {\nonnegative} and measurable (see
Section~\ref{sec:MeasSect}), one can take their integral.
The candidate product measure is the function defined on the product
$\sigma$-algebra~$\Sigma_1\otimes\Sigma_2$ (see Section~\ref{sec:ProdSigAlg})
by
\begin{equation}
  \label{eq:prod-meas}
  (\mu_1 \otimes \mu_2) (A) := \int_{X_1} \mu_2 (s_{x_1} (A)) \, d\mu_1,
\end{equation}
again represented in {\Coq} by a total function,
\begin{lstlisting}
Definition meas_prod_meas : (X1 * X2 -> Prop) -> Rbar :=
  fun A => LInt_p muX1 (meas_section muX2 A).
\end{lstlisting}

We easily deduce that this candidate function is both {\nonnegative} and equal
to zero on the empty set.
The $\sigma$-additivity property is obtained by means of $\sigma$-additivity of
the integral (see Section~\ref{sec:LInt_p}), and of the measure~$\mu_2$.
This proves that the candidate is a measure, and that we can instantiate the
record defining the product measure \coqe{meas_prod} as an object of type
measure (see Section~\ref{sec:Meas}), so all the proved results on measures are
available.

Moreover, Equation~\eqref{eq:box-meas-section}, and the positive linearity of
the integral ensure the box property~\eqref{eq:ProdMeasBox}, thus making
\coqe{meas_prod} a product measure.

\myskip

Product measures are proved to keep the finiteness, or $\sigma$-finiteness,
property of the initial measures~$\mu_1$ and~$\mu_2$: for all measure~$\mu$ on
$(X_1\times X_2,\Sigma_1\otimes\Sigma_2)$ satisfying the box
property~\eqref{eq:ProdMeasBox}, we have~$\mu_1$ and~$\mu_2$ finite
$\Implies\mu$ finite, and~$\mu_1$ and~$\mu_2$ $\sigma$-finite $\Implies\mu$
$\sigma$-finite.

Then, the proof of uniqueness of the product measure follows exactly the same
path as the one for the measurability of measure of sections (see
Section~\ref{sec:MeasSect}).
Firstly, when the measures~$\mu_1$ and~$\mu_2$ are finite, we introduce two
(finite) product measures~$m$ and~$\tm$ induced by~$\mu_1$ and~$\mu_2$ ({\ie}
both satisfying~\eqref{eq:ProdMeasBox}).
The set $\calS\eqdef\{A\in\Sigma_1\otimes\Sigma_2\st m(A)=\tm(A)\}$ is proved
to contain~$\Sigma_1\otimes\Sigma_2$ using \coqe{monotone_class_Prop}, which
shows uniqueness.
Then, the result is again extended to $\sigma$-finite measures by means of
restricted measures.

\section{The {\Ton} Theorem}
\label{sec:Ton}

With the product measure built in Section~\ref{sec:Prod}, we can now consider
the integration of {\nonnegative} measurable functions on a product space.
As in Section~\ref{sec:Prod}, we assume that the measures are $\sigma$-finite,
which ensures existence and uniqueness of the product measure.

More precisely, this section deals with the proof of the {\TonelliTh} that
allows to compute a double integral on a product space by integrating
successively with respect to each variable, either way.
Besides the following formulas, the theorem also states measurability
properties that ensure legitimacy of all integrals (see
Theorem~\ref{th:tonelli}):
\begin{align}
  \label{eq:tonelli-eq1}
  \int_{X_1 \times X_2} f (x_1, x_2) \, d(\mu_1 \otimes \mu_2) (x_1, x_2)
  &= \int_{X_1} \left(
    \int_{X_2} f (x_1, x_2) \, d\mu_2 (x_2) \right) \, d\mu_1 (x_1)\\
  \label{eq:tonelli-eq2}
  &= \int_{X_2} \left(
    \int_{X_1} f (x_1, x_2) \, d\mu_1 (x_1) \right) \, d\mu_2 (x_2).
\end{align}

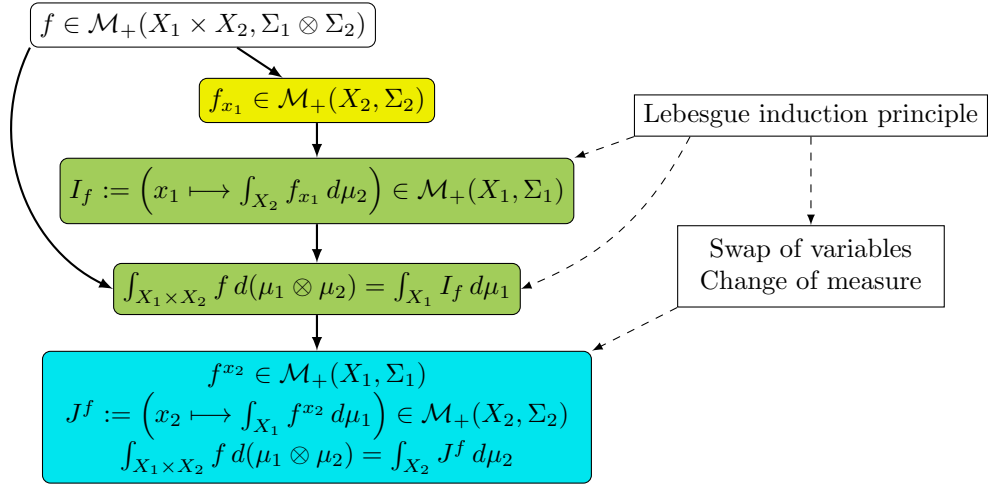
\begin{figure}[htb]
\begin{center}
\begin{tikzpicture}
\node[draw,rounded corners=3pt] (f) at (2,6.2)
  {$f\in\calMplus(X_1\times X_2,\Sigma_1\otimes\Sigma_2)$};
\node[draw,rounded corners=3pt,fill=\colorB] (Secf) at (3.5,5.2)
  {$f_{x_1}\in\calMplus(X_2,\Sigma_2)$};
\node[draw,rounded corners=3pt,fill=\colorC] (If) at (3.5,4)
  {$I_f:=\left(x_1\longmapsto\int_{X_2}f_{x_1}\,d\mu_2\right)
      \in\calMplus(X_1,\Sigma_1)$};
\node[draw,rounded corners=3pt,fill=\colorC] (T1) at (3.5,2.7)
  {$\int_{X_1\times X_2}f\,d(\mu_1\otimes\mu_2)=\int_{X_1}I_f\,d\mu_1$};
\node[draw,rounded corners=3pt,fill=\colorD] (T2) at (3.5,1)
  {$\begin{array}{c}
      f^{x_2}\in\calMplus(X_1,\Sigma_1)\\
      J^f:=\left(x_2\longmapsto\int_{X_1}f^{x_2}\,d\mu_1\right)
      \in\calMplus(X_2,\Sigma_2)\\
      \int_{X_1\times X_2}f\,d(\mu_1\otimes\mu_2)=\int_{X_2}J^f\,d\mu_2
    \end{array}$};
\draw[->,>=latex,thick] (f) to (Secf);
\draw[->,>=latex,thick] (Secf) to (If);
\draw[->,>=latex,thick] (If) to (T1);
\draw[->,>=latex,thick] (T1) to (T2);
\draw[->,>=latex,thick] (f.south west) to [bend left=-45] (T1.west);
\node[draw] (LCS) at (10,5) {\LIP};
\node[draw] (Swap) at (10,3)
  {$\begin{array}{c}
      \mbox{Swap of variables}\\
      \mbox{Change of measure}
    \end{array}$};
\draw[->,>=latex,dashed] (LCS.south west) to (If.north east);
\draw[->,>=latex,dashed] (LCS.190) to [bend right=-20] (T1.east);
\draw[->,>=latex,dashed] (LCS) to (Swap);
\draw[->,>=latex,dashed] (Swap.south west) to (T2.north east);
\end{tikzpicture}
\end{center}
\caption{%
  Flowchart illustrating the construction of the iterated integrals on a
  product space.\protect\\
  The fill colors refer to sections:
  \ref{sec:SecFun} in {\secBColorName},
  \ref{sec:TonelliTh1} in {\secCColorName}, and
  \ref{sec:TonelliTh2} in {\secDColorName}.\protect\\
  Dashed lines denote the use of the listed proof arguments, that were
  developed for the present work.}
\label{fig:double-int}
\end{figure}

Similarly to the process used in Section~\ref{sec:Prod}, the iterated integral
(right-hand side of~\eqref{eq:tonelli-eq1}) is built in three steps, see
Figure~\ref{fig:double-int}.
Firstly, $X_1$-sections of functions are proved to be $\Sigma_2$-measurable.
Then, the integral (in~$X_2$) of sections of functions is proved to be
$\Sigma_1$-measurable.
And the iterated integral is the integral (in~$X_1$) of the integral (in~$X_2$)
of the sections of functions.
Finally, Formula~\eqref{eq:tonelli-eq1} is first proved, and
then~\eqref{eq:tonelli-eq2} is deduced from the latter by a swap of variables
relying both on a change of measure and on the uniqueness of the product
measure.

The main argument for this proof is the {\LIP} (see Section~\ref{sec:LIP}).
It is used twice: to obtain the measurability of the integral of sections of
functions together with the first Tonelli formula, and for the
change-of-measure formula for the integral.

Section~\ref{sec:SecFun} is dedicated to sections of functions, and
Section~\ref{sec:TonelliTh1} to the iterated integral and the proof of the first
formula of the {\TonelliTh}.
Finally, the full proof of the {\TonelliTh} is obtained in
Section~\ref{sec:TonelliTh2}.

\subsection{Section of Function}
\label{sec:SecFun}

Similarly to sections of subsets in Section~\ref{sec:Sect}, given a numeric
function~$f:X_1\times X_2\to\Rbar$ and{\mypointb}~$x_1\in X_1$, the
\emph{$X_1$-section of~$f$ at~$x_1$} is the partial application
$f_{x_1}:=(x_2\mapsto f(x_1,x_2))$.
\begin{lstlisting}
Definition section_fun : X1 -> (X1 * X2 -> Rbar) -> X2 -> Rbar := fun x1 f x2 => f (x1, x2).
\end{lstlisting}

From measurability of sections of subsets, we deduce that, if~$f$ is
in~$\calMplus(X_1\times X_2,\Sigma_1\otimes\Sigma_2)$, then its $X_1$-sections
are in $\calMplus(X_2,\Sigma_2)$ (the {\nonnegativity} property is obvious).
\begin{lstlisting}
Lemma section_fun_Mplus : \forall f x1, Mplus genX1xX2 f -> Mplus genX2 (section_fun x1 f).
\end{lstlisting}

Symmetrically, for any~$x_2\in X_2$, we introduce the
\emph{$X_2$-section of~$f$ at~$x_2$}, the partial application with respect to
the second variable, $f^{x_2}:=(x_1\mapsto f(x_1,x_2))$.

\subsection{Iterated Integral and the First Formula of the {\Ton} Theorem}
\label{sec:TonelliTh1}

As sections of functions are {\nonnegative} and $\Sigma_2$-measurable (see
Section~\ref{sec:SecFun}), one can take their integral (in~$X_2$).
For any function $f\in\calMplus(X_1\times X_2,\Sigma_1\otimes\Sigma_2)$, we
define
\begin{equation*}
  I_f := \left( x_1 \longmapsto \int_{X_2} f_{x_1} \, d\mu_2 \right).
\end{equation*}
\begin{lstlisting}
Definition LInt_p_section_fun : (X1 * X2 -> Rbar) -> X1 -> Rbar :=
  fun f x1 => LInt_p muX2 (section_fun x1 f).
\end{lstlisting}

The iterated integral corresponds to integrate once more (in~$X_1$), but one
must first establish that~$I_f\in\calMplus(X_1,\Sigma_1)$.
The {\nonnegativity} result directly follows from the {\monotony} of the
integral (see Section~\ref{sec:LInt_p}).
The general measurability result, together with the first Tonelli
formula~\eqref{eq:tonelli-eq1}, are proved by means of the {\LIP} of
Section~\ref{sec:LIP}.

\myskip

Let us first review the properties of the function $I:=(f\mapsto I_f)$.
From the properties of the integral, $I$~is monotone and positive linear.
In the case of indicator functions, for any~$x_1\in X_1$, the section reads
$(\charac{A})_{x_1}=\charac{s_{x_1}(A)}$, which yields the formula
$I_{\charac{A}}(x_1)=\mu_2(s_{x_1}(A))$.
And from the {\BeppoLeviMonotConvTh} (see Section~\ref{sec:LInt_p}),
$I$~commutes with the supremum:
for all {\nondecreasing} sequence~$(f_n)_{n\in\N}$
in~$\calMplus(X_1\times X_2,\Sigma_1\otimes\Sigma_2)$, we have the equality
\mydisplaymaths{I_{\sup_{n\in\N}f_n}=\sup_{n\in\N}I_{f_n}}

\myskip

Let~\coqe{P0 f := Mplus genX1 (LInt_p_section_fun f)} be the predicate of the
{\nonnegativity} and measurability of~$I_f$, of type
\coqe{(E -> Rbar) -> Prop}.
Then, previous formulas and closedness properties of~$\calMplus$ (see
Section~\ref{sec:MeasFun}) provide the compatibility of~\coqe{P0} with
indicator functions, positive linearity, and the supremum of {\nondecreasing}
sequences.
For instance, we have
\begin{lstlisting}
Lemma LInt_p_section_fun_measurable_plus :
  \forall f g, Mplus genX1xX2 f -> Mplus genX1xX2 g ->
    P0 f -> P0 g -> P0 (fun x => Rbar_plus (f x) (g x)).
\end{lstlisting}

\myskip

Let us now define the predicate~\coqe{P} of the existence of the iterated
integral (granted by~\coqe{P0}) and the validity of the first Tonelli formula
of~\eqref{eq:tonelli-eq1}:
\begin{lstlisting}
Let P : (E -> Rbar) -> Prop :=
  fun f => P0 f /\ LInt_p meas_prod f = LInt_p muX1 (LInt_p_section_fun f).
\end{lstlisting}
where \coqe{meas_prod} is the product measure defined in
Section~\ref{sec:ProdMeas}.
Again, the compatibility of~\coqe{P} with indicator functions, positive
linearity, and the supremum is easily obtained from the previous results.
Namely, we have
\begin{lstlisting}
Lemma LInt_p_section_fun_meas_prod_charac :
  \forall A, measurable genX1xX2 A -> P (charac A).

Lemma LInt_p_section_fun_meas_prod_scal :
  \forall a f, 0 <== a -> Mplus genX1xX2 f -> P f -> P (fun x => Rbar_mult a (f x)).

Lemma LInt_p_section_fun_meas_prod_plus :
  \forall f g, Mplus genX1xX2 f -> Mplus genX1xX2 g -> P f -> P g -> P (fun x => Rbar_plus (f x) (g x)).

Lemma LInt_p_section_fun_meas_prod_Sup_seq :
  \forall f, incr_fun_seq f -> Mplus_seq genX1xX2 f ->
    (\forall n, P (f n)) -> P (fun x => Sup_seq (fun n => f n x)).
\end{lstlisting}

\myskip

Now, the first part of the {\TonelliTh} can be stated in {\Coq} as
\begin{lstlisting}
Lemma Tonelli_aux1 :
  \forall f, Mplus genX1xX2 f ->
    Mplus genX1 (LInt_p_section_fun f) /\
    LInt_p meas_prod f = LInt_p muX1 (LInt_p_section_fun f).
\end{lstlisting}
And its proof is a direct application of the {\LIP} (see Section~\ref{sec:LIP})
with the predicate~\coqe{P}, as all the premises corresponds to the previous
lemmas.

\subsection{Change of Measure, Second Formula, and the {\Ton} Theorem}
\label{sec:TonelliTh2}

There is no doubt that the second formula~\eqref{eq:tonelli-eq2} can be proved
using the same path as the first claim: use sections with respect to the second
variable, define~$J^f$ (see Figure~\ref{fig:double-int}),
prove~$J^f\in\calMplus$ and the equality by the {\LIP}.
This would be easy, but pretty long and redundant.
Instead, we have exploited the ``symmetry'' between the right-hand sides of
both formulas.
The first idea is a simple exchange of the roles of the two variables that
expresses the previous result for functions of type \coqe{X2 * X1 -> Rbar}.
And then, the difficult part is a change of measure that brings back to the
target type \coqe{X1 * X2 -> Rbar}.

In the framework of the Lebesgue integral, the change of measure is an
application of the concept of \emph{image measure} ({\eg}
see~\cite{mai:m2:14}), also called \emph{pushforward measure} as the measure is
transported between $\sigma$-algebras, here from~$\Sigma_2\otimes\Sigma_1$
to~$\Sigma_1\otimes\Sigma_2$.

\subsubsection{Change of measure}
\label{sec:ChgMeas}

Let~$(X,\Sigma)$ and~$(Y,\calT)$ be measurable spaces.
Let~$h:X\to Y$~be a function and~\coqe{Mh} be a proof of its measurability.
Let~$\mu$ be a measure on~$(X,\Sigma)$.
The \emph{image measure of~$\mu$ by~$h$} is the measure on~$(Y,\calT)$ defined
by~$h \imgmeas \mu:=\mu\circ h^{-1}$, and denoted in {\Coq} by
\coqe{meas_image h Mh mu}.
The proof that it is indeed a measure directly follows from the measure
properties of~$\mu$, and~\coqe{Mh}.

Now, given $g\in\calMplus(Y,\calT)$, the compatibility of measurability with
the composition of functions provides $g\circ h\in\calMplus(X,\Sigma)$, and one
has the following change-of-measure formula,
\begin{equation}
  \label{eq:change-meas}
  \int_Y g \, d(h \imgmeas \mu) = \int_X g \circ h \, d\mu.
\end{equation}
\begin{lstlisting}
Lemma LInt_p_change_meas :
  \forall g, Mplus genY g -> LInt_p (meas_image h Mh mu) g = LInt_p mu (fun x => g (h x)).
\end{lstlisting}
The proof follows the {\LIP} with the predicate~\coqe{P'} corresponding
to~\eqref{eq:change-meas}.
Once again, the compatibility of~\coqe{P'} with indicator functions, positive
linearity, and the supremum directly follows from properties of the integral,
such as positive linearity and the {\BeppoLeviMonotConvTh}.

\subsubsection{Swap and Second Formula}
\label{sec:Swap}

Using Section~\ref{sec:ProdMeas}, let~$\mu_{12}:=\mu_1\otimes\mu_2$ be the
product measure induced by~$\mu_1$ and~$\mu_2$ on the product
space~$(X_1\times X_2,\Sigma_1\otimes\Sigma_2)$.
In {\Coq}, \coqe{muX1xX2 := meas_prod muX1 muX2}.
By exchanging the two spaces, let~$\mu_{21}:=\mu_2\otimes\mu_1$ be the product
measure induced by~$\mu_2$ and~$\mu_1$
on~$(X,\Sigma):=(X_2\times X_1,\Sigma_2\otimes\Sigma_1)$.
In {\Coq}, \coqe{muX2xX1 := meas_prod muX2 muX1}.

Let~$h:(x_2,x_1)\in X_2\times X_1\mapsto(x_1,x_2)\in X_1\times X_2$ be the swap
of variables.
We construct the proof~\coqe{Mh} of its measurability.
The image measure~$h \imgmeas \mu_{21}$ is defined on the measurable
space $(Y,\calT):=(X_1\times X_2,\Sigma_1\otimes\Sigma_2)$.
In {\Coq}, \coqe{meas_prod_swap := meas_image h Mh muX2xX1}.
The proof that it is a product measure induced by~$\mu_1$ and~$\mu_2$ is
straightforward.

Now, let $f\in\calMplus(X_1\times X_2,\Sigma_1\otimes\Sigma_2)$.
One has~$f\circ h\in\calMplus(X_2\times X_1,\Sigma_2\otimes\Sigma_1)$, and
using the section with respect to the second variable (see
Section~\ref{sec:SecFun}), we have
\begin{equation}
  \label{eq:tonelli-blop}
  \forall x_2 \in X_2,\quad
  f^{x_2}
  := (x_1 \longmapsto f (x_1, x_2))
  = (x_1 \longmapsto f \circ h (x_2, x_1))
  = (f \circ h)_{x_2}.
\end{equation}

We then deduce the second part of the {\TonelliTh}~\eqref{eq:tonelli-eq2} from
the previous ingredients:
\begin{align*}
  \int_{X_1 \times X_2} f \, d\mu_{12}
  &\stackrel{(a)}{=} \int_{X_1 \times X_2} f \, d(h \imgmeas \mu_{21})
  \stackrel{(b)}{=} \int_{X_2 \times X_1} f \circ h \, d\mu_{21}\\
  &\stackrel{(c)}{=} \int_{X_2} \left( \int_{X_1} (f \circ h)_{x_2} \, d\mu_1 \right) \, d\mu_2
  \stackrel{(d)}{=} \int_{X_2} \left( \int_{X_1} f^{x_2} \, d\mu_1 \right) \, d\mu_2.
\end{align*}
Uniqueness of the product measure of Section~\ref{sec:ProdMeas}
yields~$h\imgmeas\mu_{21}=\mu_{12}$, and thus gives~(a).
The above change-of-measure formula~\eqref{eq:change-meas} gives~(b).
The first formula of the {\TonelliTh}~\eqref{eq:tonelli-eq1} applied
to~$X_2 \times X_1$ gives~(c).
The above Equation~\eqref{eq:tonelli-blop} gives~(d).

This second part of {\TonelliTh} can be stated in {\Coq} as
\begin{lstlisting}
Lemma Tonelli_aux2 :
  \forall f, Mplus genX1xX2 f ->
  Mplus genX2 (LInt_p_section_fun muX1 (swap f)) /\
  LInt_p meas_prod_swap f = LInt_p muX2 (LInt_p_section_fun muX1 (swap f)).
\end{lstlisting}
where \coqe{swap f} denotes~$f\circ h$.

\subsubsection{Statement of the {\Ton} Theorem}
\label{sec:ProofTon}

Finally, we formalize the {\TonelliTh} that gathers the two
equalities~\eqref{eq:tonelli-eq1} and~\eqref{eq:tonelli-eq2}.
We assume that~$X_1$ and~$X_2$ are {\nonempty} and that~$\mu_1$ and~$\mu_2$ are
$\sigma$-finite measures.
Then,
\begin{lstlisting}
Lemma Tonelli_formulas :
  \forall f, Mplus genX1xX2 f ->
    LInt_p muX1xX2 f = LInt_p muX1 (LInt_p_section_fun muX2 f) /\
    LInt_p muX1xX2 f = LInt_p muX2 (LInt_p_section_fun muX1 (swap f)).
\end{lstlisting}
where \coqe{muX1xX2} stands for the product measure.
We also provide a more comprehensive but less readable theorem \coqe{Tonelli}
that moreover ensures the legitimacy of all integrals.

\section{Conclusion and perspectives}
\label{sec:Concl}

In this paper, we present the formalization and the construction of the full
formal proof of the {\TonelliTh}.
We have constructed the product measure of two $\sigma$-finite measures, built
the two iterated integrals, and proved they are equal to the double integral on
the product measure space.
A key point is the definition of {\nonnegative} measurable functions as an
inductive type.
It has been proved equivalent to the common mathematical definition and has led
to a very useful induction scheme.
Although the induction principle is present in some formalizations, building it
from an inductive type is an original point of view we have not seen in the
literature.

To achieve the proof of the {\TonelliTh}, we have also formalized in {\Coq}
common generic results and constructions such as the {\MonotClassTh}, the
restricted measure, the image measure, and a change-of-measure formula for the
integral.
The latter, combined with a swap of variables, has prevented redundancies in
our proofs.

This work confirms the fact that the library we are developing, in line with
the choices of the {\Coquelicot} library, is rather comprehensive and
usable.
First, this work has led to few additions in the core of the library, except
for the inductive definition for~$\calMplus$ that is related to the needed
{\LIP}.
Second, the library seems easy to learn.
One co-author of this article and this {\Coq} development was a novice who did
not actually participate in the previous developments.

\myskip

The natural extension after the {\TonelliTh} on {\nonnegative} measurable
functions is the {\FubiniTh} that provides the same formulas for integrable
functions with arbitrary sign.
But we would rather directly consider the version using the Bochner
integral~\cite{BCL22} that applies to functions taking their values in a Banach
space, such as the Euclidean spaces~$\R^n$ and the Hermitian spaces~$\C^n$.
For that, we can take inspiration from the work by van~Doorn in
{\Lean}~\cite{VanDoo21}, and in particular with the concept of ``marginal
integral'' that seems to be an elegant way to handle integrals on a finitary
Cartesian product.

Our long-term purpose is to formally prove the correctness of parts of a
library implementing the Finite Element Method (FEM), which is used to compute
approximated solutions of Partial Differential Equations (PDEs).
We already formalized the {\LaxMilgramTh}~\cite{BCF21}, one of the key
ingredient to numerically solve PDEs, and we need to build suitable Hilbert
functional spaces on which to apply it.
The target candidates are the Sobolev spaces such as~$H^1$, that represents
square integrable functions with square integrable first derivatives.
Of course, this will involve the formalization of the~$L^p$ Lebesgue spaces as
complete normed vector spaces, and parts of the distribution
theory~\cite{sch:td:66}.

\bibliographystyle{plainnat}
\bibliography{biblio}

\end{document}